\documentclass{emulateapj}
\shorttitle{Gini Coefficients in Strongly Lensed Galaxies}
\shortauthors{Florian et al.}

\begin{document}

\title{The Gini Coefficient as a Morphological Measurement of Strongly Lensed Galaxies in the Image Plane}

\author{Michael K. Florian\altaffilmark{1,2}}

\author{Nan Li\altaffilmark{1,2,3}}

\author{Michael D. Gladders\altaffilmark{1,2}}

\altaffiltext{1}{Department of Astronomy and Astrophysics, University of Chicago, Chicago, IL 60637}
\altaffiltext{2}{Kavli Institute for Cosmological Physics, The University of Chicago, Chicago, IL 60637}
\altaffiltext{3}{Argonne National Laboratory, 9700 South Cass Avenue B109, Lemont, IL 60439}

\begin{abstract}
Characterization of the morphology of strongly lensed galaxies is challenging because images of such galaxies are typically highly distorted.  Lens modeling and source plane reconstruction is one approach that can provide reasonably undistorted images from which morphological measurements can be made, although at the expense of a highly spatially variable telescope PSF when mapped back to the source plane.  Unfortunately, modeling the lensing mass is a time and resource intensive process, and in many cases there are too few constraints to precisely model the lensing mass.  If, however, useful morphological measurements could be made in the image plane rather than the source plane, it would bypass this issue and obviate the need for a source reconstruction process for some applications.  We examine the use of the Gini coefficient as one such measurement.  Because it depends on the cumulative distribution of the light of a galaxy, but not the relative spatial positions, the fact that surface brightness is conserved by lensing means that the Gini coefficient may be well-preserved by strong gravitational lensing.  Through simulations, we test the extent to which the Gini coefficient is conserved, including by effects due to PSF convolution and pixelization, to determine whether it is invariant enough under lensing to be used as a measurement of galaxy morphology that can be made in the image plane.
\end{abstract}

\keywords{gravitational lensing: strong --- galaxies: formation ---- galaxies: evolution --- galaxies: high-redshift --- methods: observational}

\section{Introduction}
Studying galaxies over the course of cosmic time requires making many measurements--measurements of broadband colors, spectra, emission and absorption lines and the implied chemical abundances, star formation rates, stellar populations, and more.  Among the observables of interest is, of course, morphology, which is of particular relevance because of its possible relationships with both galaxy and star formation (e.g., \citealp{Dress80,Ab1996,Will15}).

While morphologies of low redshift galaxies can be studied in great detail, high redshift galaxies present a challenge because of their small angular sizes.  Even with the increased signal to noise ratios (S/N) and resolutions of the James Webb Space Telescope (JWST) and the Wide-Field Infrared Survey Telescope (WFIRST), this will continue to be a limitation at large redshifts.  A desire to study small-scale structures in high redshift galaxies has led some to take advantage of the extra magnification provided by strong gravitational lensing (e.g., \citealp{Wuyts14,Swin09,Jones10,Jones13}).

Much effort has been put into quantifying the morphological properties of unlensed galaxies.  Current morphological metrics include the CAS system which consists of the concentration parameter, the asymmetry parameter and the clumpiness parameter as described by \citet{Con2003} and references therein, as well as the Gini coefficient \citep{Ab2003}, and the internal color dispersion \citep{Pap2003}, among others.  However, the definitions of many of these measurements depend on the relative amplitude \textit{and} position of a galaxy's light, and so are unlikely to be useful when applied directly to strongly lensed images of distant galaxies.  Given a lens model, such images can be mapped back to the undistorted source plane, but creating models that allow for the reconstruction of the source galaxy can be a difficult and time consuming process and even when mass models can be constructed with low levels of uncertainty, uncertainties in magnification maps can still be quite high.  Moreover, simple source plane reconstruction techniques, absent some form of point spread function (PSF) deconvolution, will also map a telescope PSF that is reasonably invariant in the image plane into a highly distorted and variable PSF in the source plane, potentially complicating any morphological analysis (though it should be noted that in systems with multiple images, the ability to make multiple measurements of the same small scale regions of a source galaxy where the telescope PSF is mapped onto each image differently, can provide some ability to work around this complication, but it still requires significant astronomer time to do so).  And, while many techniques exist for modeling lenses (see \citealt{Lef13} for a review), in some cases there are simply not enough constraints to make a lens model possible at all without major uncertainties.  When sufficient constraints are available, it can still take many hours of focused effort for a researcher or group to produce a final model.  This modeling, required before reconstruction of the undistorted source galaxy can be attempted, represents a major bottleneck in the process of making morphological measurements of strongly lensed galaxies.  This difficulty will only become more severe as the next generation of ground- and space-based survey telescopes comes online and brings our samples of strong lensing systems into the thousands.

To avoid this difficulty altogether, it is worth seeking out morphological measurements that can be performed in the image (lensed) plane rather than the source (unlensed) plane.  In this paper, we show through simulation that the Gini coefficient, introduced to astronomy and used for classifying galaxy morphologies in the Sloan Digital Sky Survey (SDSS) by \citet{Ab2003}, is one such measurement.  With at least one metric available for use in both lensed and unlensed galaxies, we can begin to contextualize the sample of strongly lensed galaxies to see if certain types of galaxies are preferentially observed in lensing (e.g., perhaps clumpy objects may be observed more often because the many high surface brightness clumps allow a higher likelihood of a bright region falling on a caustic than in objects where the light is more concentrated).  In addition, folding morphological measurements into existing methods for identifying image families in strong lensing systems may help our ability to correctly and quickly identify multiple images of a single source galaxy.  And, if proxies for or redefinitions of other morphological measurements can be found for strongly lensed sources, it may be possible to adapt or reformulate morphological classification systems like the CAS system described above or the Gini-M$_{20}$ system \citep{Lotz04} to allow for usage in the case of strongly lensed sources.  This would be particularly useful because while new generations of telescopes will increase our ability to access small-scale structures in unlensed galaxies at higher and higher redshits, the smallest accessible spatial scales at the highest observable redshifts will always be a result of gravitational lensing.  Therefore, detailed studies of the evolution of galaxy morphologies at the earliest times will rely on being able to measure morphological characteristics of strongly lensed objects.

In this paper, we show that a relationship exists between lensed and unlensed Gini coefficients for a given source image with a resolution similar to images taken by the Hubble Space Telescope (HST) as long as the apertures are carefully defined.

\section{Testing the Stability of the Gini Coefficient: The Simulations}
The Gini coefficient, as applied to galaxy morphologies, is a measurement of the inequality of the distribution of light in a galaxy.  Conceptually, this measurement is made by ordering the pixels that make up the image of a galaxy in ascending order by flux and then comparing the resulting cumulative distribution function to what would be expected from a perfectly even flux distribution.   A low Gini coefficient (close to zero) means that the distribution of light is fairly uniform, while a high Gini coefficient (close to one) means that most of the galaxy's light is contained in only a small fraction of the pixels.  In practice, the Gini coefficient is calculated using the following formula from \citet{Lotz04}:
\begin{equation}
G = \frac{1}{|\overline{X}|n(n - 1)} \sum_{i=1}^{n} (2i - n - 1)|X_{i}|
\end{equation}
where $X_{i}$ is the value of the $i^{th}$ pixel when ordered by flux, $\overline{X}$ is the mean of the pixel values, and $n$ is the total number of pixels.  Because gravitational lensing preserves surface brightness, the cumulative distribution of the light should be little changed, and the Gini coefficient can be expected to be fairly well preserved by lensing.

This hypothesis was tested by simulating the lensing effect on actual images of low redshift galaxies treated as if they were galaxies at higher redshifts using steps which will be explained in further detail in the following sections.  Briefly stated, a set of 33 detailed images of low redshift galaxies ($z \lesssim 0.45$) were chosen from the mosaics produced by the CANDELS team \citep{Grogin11,Koek11} and selected from the CANDELS UDS catalogue \citep{CANDELS_UDS}.  Those galaxies were placed at higher redshifts and on or near caustics of a galaxy cluster scale lensing mass and run through a gravitational lensing simulation code to produce arcs of a similar size to those seen in observations.  These final images of the arcs were then convolved with an HST-like PSF and rebinned to a pixel scale of 0.03 arcseconds/pixel.  Varying amounts of noise were added to the images, and masks were made based on a generalization of the Petrosian radius defined in such a way as to be applicable to objects of arbitrary shape.  Within these apertures, Gini coefficients were calculated and compared in a variety of ways which will be detailed in section 3.

\subsection{Galaxy Selection}
The 33 galaxies that were used as sources in the lensing simulation were selected from the CANDELS UDS field \citep{CANDELS_UDS}. They were chosen to span a range of morphologies (11 ellipticals, 20 spirals, and 2 irregular galaxies).  It is difficult to know whether these galaxies have morphologies similar to galaxies at z $\approx$ 2 since small spatial scales are difficult enough to access in such galaxies that significant debate remains over their typical morphologies.  However, these galaxies do span a range of morphologies and a range of Gini coefficients ($\approx$ 0.2 to $\approx$ 0.6) similar to those observed in previous studies at low to moderate redshifts (see, for example, \citealp{Ab2003} for low redshift galaxies and \citealp{Lotz06} for galaxies at z $\approx$ 1.5 and z $\approx$ 4).  We chose galaxies that are low redshfit, and large on the sky, so that detailed images of small scale structures were available.  Objects chosen were typically between redshifts 0.2 and 0.4, corresponding to spatial scales of approximately 100--150 parsecs per pixel.  Before lensing, an aperture was made based on a 12$\sigma$ threshold on the stack of the F160W, F814W and F606W images, which, given the high signal-to-noise ratios in the CANDELS UDS images, still included most of the light associated with each galaxy.  Pixels outside of these apertures were set to zero, minimizing the number of pixels that needed to be treated by the gravitational lensing ray-tracing code while also isolating the target galaxies from nearby objects.  

\subsection{Simulating Gravitational Lensing by Ray-Tracing}
In our first gravitational lensing image simulation, the lensing mass had a spherical NFW profile with virial mass M$_{200}$ = 10$^{15}$ M$_{\odot}/h$ and concentration parameter c = 5.  The lensing mass was placed at redshift $z_{l}$ = 0.2.  The source plane was located at $z_{s}$ = 1.0.  Because they were actually observed at a redshift much closer to 0, the 33 galaxies selected from the CANDELS UDS field were treated as though each pixel was 0.0075 arcseconds in the source plane even though the original images were drizzled to a 0.03 arcsecond/pixel resolution.  For each galaxy we picked 50 random positions inside an 8 arcsecond square grid, centered on the lens so that the image positions would be close to caustics.  For images of each galaxy in each of 3 filters (F160W, F814W, and F606W), the ray-tracing was performed using the code described by \citet{Li2015} at each of the 50 source positions, resulting in 4950 total lensed images (33 galaxies $\times$ 50 positions $\times$ 3 filters).  At this stage, the images were sampled to a 0.01 arcsecond pixel grid.  It is important to note that no SED shifting was done to convert observed low-z SEDs to z $\approx$ 1 SEDs.  Simulations using the ``F160W" filter are therefore intended to portray rest frame NIR emission rather than optical or UV emission that was redshifted.  For the purpose of testing the preservation of the Gini coefficient, having a realistic spatial light distribution is all that matters.  For artificially redshifted photometry to add to our analysis, it would also have to account for morphological differences as a function of rest-frame wavelength, which is still highly unconstrained at all but the lowest redshifts.

The images produced by the ray-tracing code have a much finer pixel scale than could actually be observed with the Hubble Space Telescope.  To create images similar to what would actually be observed, these images were convolved with a Gaussian PSF with the same FWHM as the HST PSF and rebinned to a final scale of 0.03 arcseconds per pixel.  To test the effects of performing observations with different S/N per pixel, noise was added to each of the resultant arc images.  For each arc, an image was produced with S/N per pixel of 10$^{-3.5}$ through 10$^{1}$ in logarithmic steps of 10$^{0.5}$, yielding 10 images with different average S/N per pixel for each arc.  These logarithmic bins were chosen to correspond encompass the range of S/N levels typical of observations performed with HST.  Thus, in summary, our database of simulated strongly lensed images includes nearly 50,000 individual frames (33 galaxies $\times$ 50 positions $\times$ 3 filters $\times$ 10 S/N levels) and many more individual arcs due to the high occurrence of systems with multiple images.

\subsection{Generalizing the Petrosian Radius to Isolate Arcs}
\citet{Lisk2008} found that the size of the aperture inside which the Gini coefficient is calculated can have a significant effect on the measured value of the Gini coefficient.  Use too small of an aperture and only part of the galaxy is used for the measurement.  Use too large of an aperture and so much sky is included that the Gini coefficient will be biased toward values near 1.  \citet{Lisk2008} measured Gini coefficients using elliptical Petrosian apertures of various sizes (i.e., using apertures defined using multiples of the Petrosian semimajor axis, which is used in place of a Petrosian radius).  Apertures that best balanced inclusion of galactic light with exclusion of sky and therefore maximized the differences between Gini coefficients of different objects were found to be constructed from semimajor axes that fell between 2/3 and 1 times the Petrosian semimajor axis.

In light of these results, it is important to define a similarly inclusive aperture for arcs that also avoids including too many sky pixels.  To do this, the definition of the Petrosian radius used by the Sloan Digital Sky Survey \citep{Blant2001,Yas2001} was adopted and then reformulated in terms of areas (rather than radii) in order to find apertures for galaxies whose shapes were severely distorted by strong gravitational lensing.

The Petrosian radius as used by SDSS is defined implicitly by the following equation from \citet{Yas2001}:
\begin{equation}
\eta = \frac{2\pi \int_{0.8r_{p}}^{1.25r_{p}}I(r)rdr/\{\pi[(1.25r_{p})^{2}-(0.8r_{p})^{2}]\}}{2\pi \int_{0}^{r_{p}}I(r)rdr/(\pi r_{p}^{2})}
\end{equation}
where $\eta$ = 0.2, $I(r)$ is the surface brightness profile of the galaxy, $r$ is the radius over which we are integrating, and $r_{p}$ is the Petrosian radius.  Essentially, this compares the average surface brightness within an annulus with an inner radius of 0.8$r$ and outer radius of 1.25$r$ to the average surface brightness in the circle of radius $r$.  When the ratio of these two values (the Petrosian ratio) is 0.2, $r$ is the Petrosian radius.

\begin{figure}
\plotone{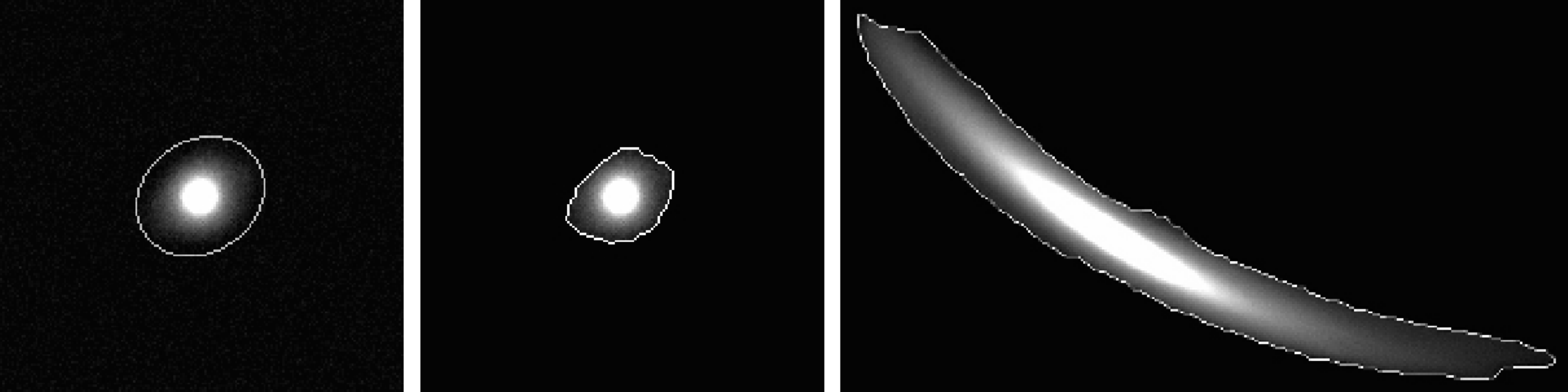}
\caption{\textit{Left:} An elliptical galaxy and its Petrosian ellipse as determined by Source Extractor. \textit{Center: } The same galaxy with its aperture determined using the method from this paper. \textit{Right: } The same galaxy after being lensed and masked using the method from this paper.}
    \label{apertures}

\end{figure}

This understanding of the Petrosian ratio leads very naturally to a redefinition in terms of areas.  We look for an annulus with an area equal to $[2\pi (1.25r)^{2} - 2\pi (0.8r)^{2}]/[2\pi r^{2}]$ or simply (1.25$^{2}$ - 0.8$^{2}$) times the area of the circle of radius $r$.  The average surface brightness within the annulus is then compared to the average surface brightness within the full circle.  When the ratio is 0.2, $r$ is the Petrosian radius.  With this reformulation in terms of areas, we extend the definition of a Petrosian aperture to include any arbitrary (connected) shape.  To do this, we take a shape, calculate its area in pixels, and build contours either inward or outward until the area of the contours contains (1.25$^{2}$ - 0.8$^{2}$) times the number of pixels that the original shape did.  When the ratio of the average surface brightness of the outer contour is 0.2 times that of the inner shape, we have found an aperture analogous to the one defined by the Petrosian radius.  Defining the initial shape of the aperture, from which we build inward or outward (usually outward), requires nothing more than thresholding.  For these simulations, we used a threshold of 2.5$\sigma$ above the background.  Fig.~\ref{apertures} shows the aperture produced by this method for an unlensed elliptical galaxy, compared to the elliptical Petrosian aperture produced for the same galaxy by Source Extractor \citep{SExtractor}.  It should be noted that while the two images with apertures defined by this method have similar Gini coefficients (0.500$\pm$0.008 for the unlensed one and 0.512$\pm$0.003 for the lensed one), these differ substantially from the Gini coefficient inside the elliptical Petrosian aperture (0.649$\pm$0.007).  Comparisons of Gini coefficients between lensed and unlensed samples must therefore both use our aperture definition.

Each simulated lensed image was masked using the method described in this section.  In some cases, especially those for which the S/N per pixel was very low, defining apertures in this manner was not possible.  At extremely low S/N, the initial aperture shape determined by thresholding does not necessarily follow the light distribution of the galaxy since it is more easily influenced by the sky noise.  As a result, these starting apertures can be quite large and quickly grow past the size of the simulated image before the Petrosian ratio falls to 0.2.  Alternatively, the starting aperture could be too small (if, for example, part of the arc is very bright, but the rest is faint) and as a result, the shape of the aperture will not accurately reflect the shape of the arc and parts of the arc may not be included.  Instances like this were rare and typically occurred only at the lowest two S/N levels that we tested and for the vast majority of lensed images, there were no such problems.

\subsection{Measuring the Gini Coefficients}
Finally, for each image of each galaxy in each filter and S/N per pixel bin, the Gini coefficient and its associated uncertainty was measured according to the prescription in \citet{Ab2003} within apertures defined as in Section 2.3.  For moderate to high S/N levels, changing the threshold of the cut used to determine the shape of the aperture from 2.5$\sigma$ to 1.5$\sigma$ or to 3.5$\sigma$ resulted in changes to the measured Gini coefficient of only about 2--3\%.  Changing the threshold level slightly changes the initial shape of the aperture, but it will not have a significant effect on the final aperture size.  This is why the Gini coefficient is hardly changed.  If the Petrosian ratio is chosen to be something other than 0.2, though, the size of the final aperture could change significantly regardless of the chosen threshold level, which would have a strong effect on the measured Gini coefficient.  As we will show in Section 3.1, such a change due to adjusting the threshold by 1$\sigma$ is still small compared to the slight variation in the measured Gini coefficient between different lensing model realizations of the same source galaxy and is therefore not a significant source of uncertainty.  However, at the very lowest S/N levels, the aperture shape can vary greatly depending on the threshold chosen, but at such low S/N levels, the sky dominates the measurement of the Gini coefficient anyway and renders it useless regardless of aperture shape.

\section{Results}
\begin{figure}
\plotone{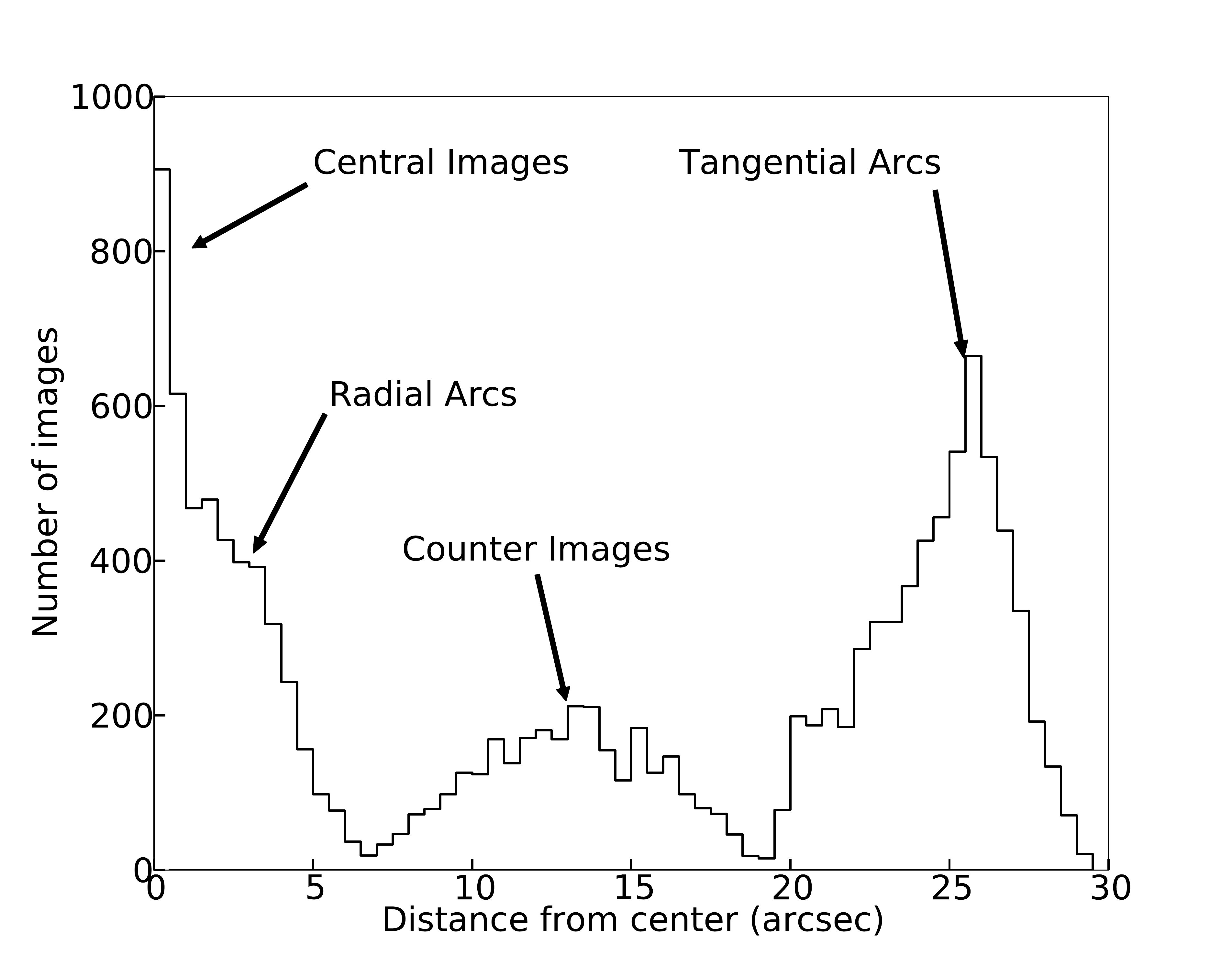}
\caption{Distribution of the distances, in arcseconds, of the nearest point on each lensed image to the center of the lensing halo.  We see four nearly distinct distributions consisting of central images, radial arcs, counter images, and tangential arcs, which allows tangential arcs and counter images to be identified using a cut based on this distance.}
    \label{locations}

\end{figure}

Before the analysis was carried out, two cuts were made on the simulated images in order to ensure that the sample was as close as possible to what would exist in an observed sample.  First, any images that were magnified by less than a factor of 4 were removed.  This mostly removed central, demagnified images that would not be observed in real data.  It also removed images where a piece of sky noise in the original unlensed CANDELS image fell on a caustic and resulted in a large image of something that wasn't actually the galaxy.  The second cut was to remove all central images and radial images so that the analysis could focus entirely on tangential arcs and counter images, which are far more likely to be observed without significant contamination from light from the intervening lens in practice.  We plot, in Fig.~\ref{locations}, the distance (in arcseconds) of the point in each of the arcs in our sample that is closest to the center of the lensing halo.  We find 4 distinct regions.  The innermost images are central images and have distances of nearly zero, which are often demagnified and unobservable.  There is another bump in the histogram centered near 3", where the radial arcs lie.  Then there are two more clearly defined peaks at around 6-18 and 20-30 arceseconds, corresponding to the counterimages and the tangential arcs, respectively.  A cut was made in this space at 6.5", removing anything closer so that only tangential arcs and counterimages were included in the sample.  All of the following results in section 3 are drawn from the sample of strongly lensed galaxy images that remained after these two cuts.

\subsection{Relationship Between Unlensed Gini Coefficient and Lensed Gini Coefficient}
We consider first the relationship between the Gini coefficient of the unlensed source galaxy and the Gini coefficient of the corresponding simulated lensed galaxy images.  We used, for the unlensed Gini coefficient, the Gini coefficient of the source galaxy as if it were placed at the same redshift as the source galaxies used in the simulation.  This required resampling the image to reduce the pixel scale by about a factor of 3 (reducing the total image size by a factor of 9).  In other words we convolved the images with a Gaussian PSF with FWHM equal to the width of the PSF of HST WFC3 or ACS images taken in the corresponding filter (either F160W, F814W, or F606W), then rebinned the resulting images from a 0.01 arcseconds/pixel effective resolution to 0.03 arcseconds/pixel and added noise as we did with the lensed images (see Section 2.2).  These unlensed images were then masked using the same method as was used for the lensed images.

\begin{figure*}
\centering
\includegraphics[width=0.75\textwidth]{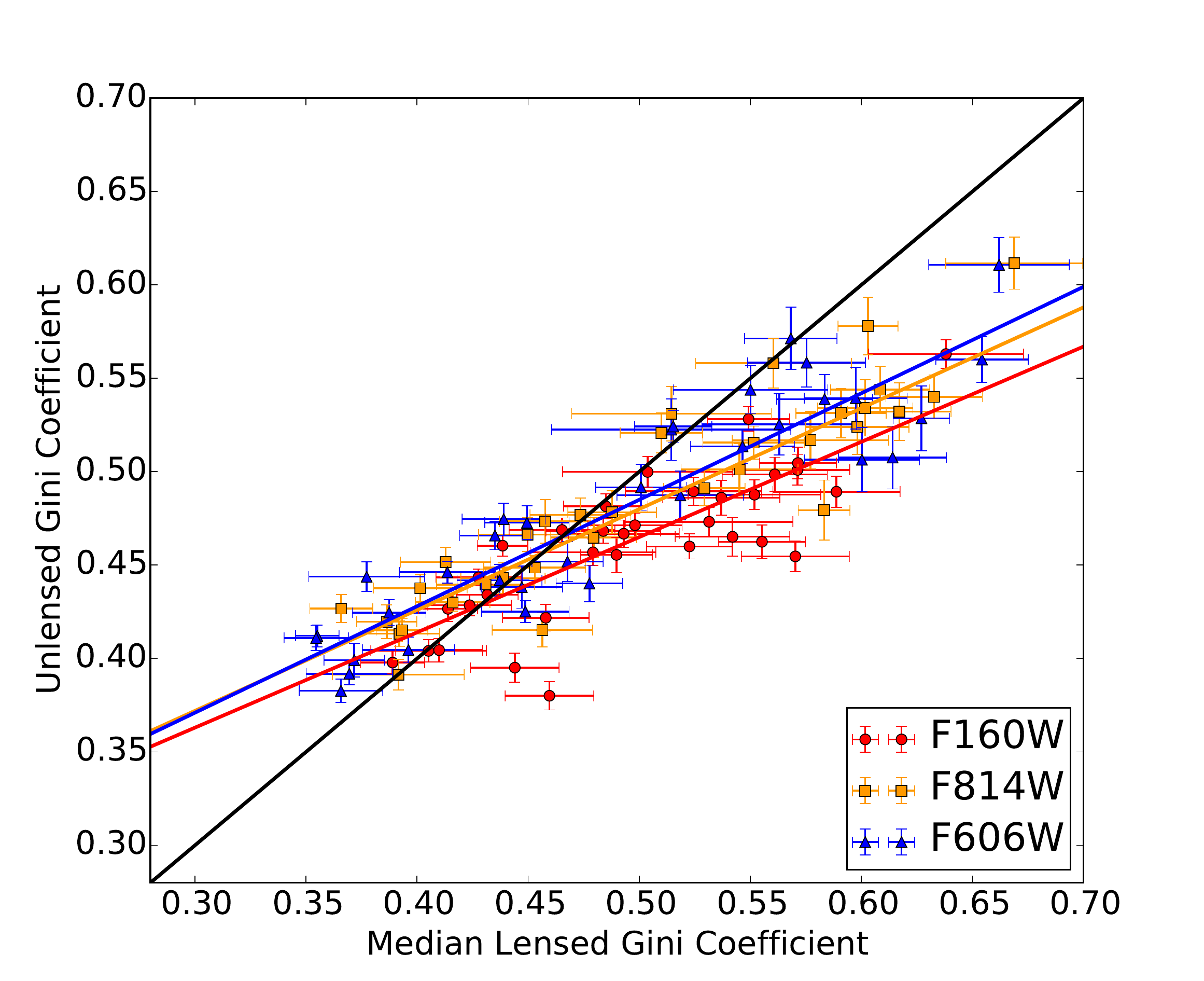}
\caption{Unlensed Gini coefficients plotted against the median lensed Gini coefficients for all filters, using arcs with average S/N per pixel of 0.1.  A relationship between unlensed and lensed gini coefficients is evident, though the exact relationship may depend on the filter (likely due to differences in PSF).  To guide the eye, the 1-to-1 line has been plotted in black.}
    \label{LGCvUGC}
\end{figure*}

\begin{figure}
\plotone{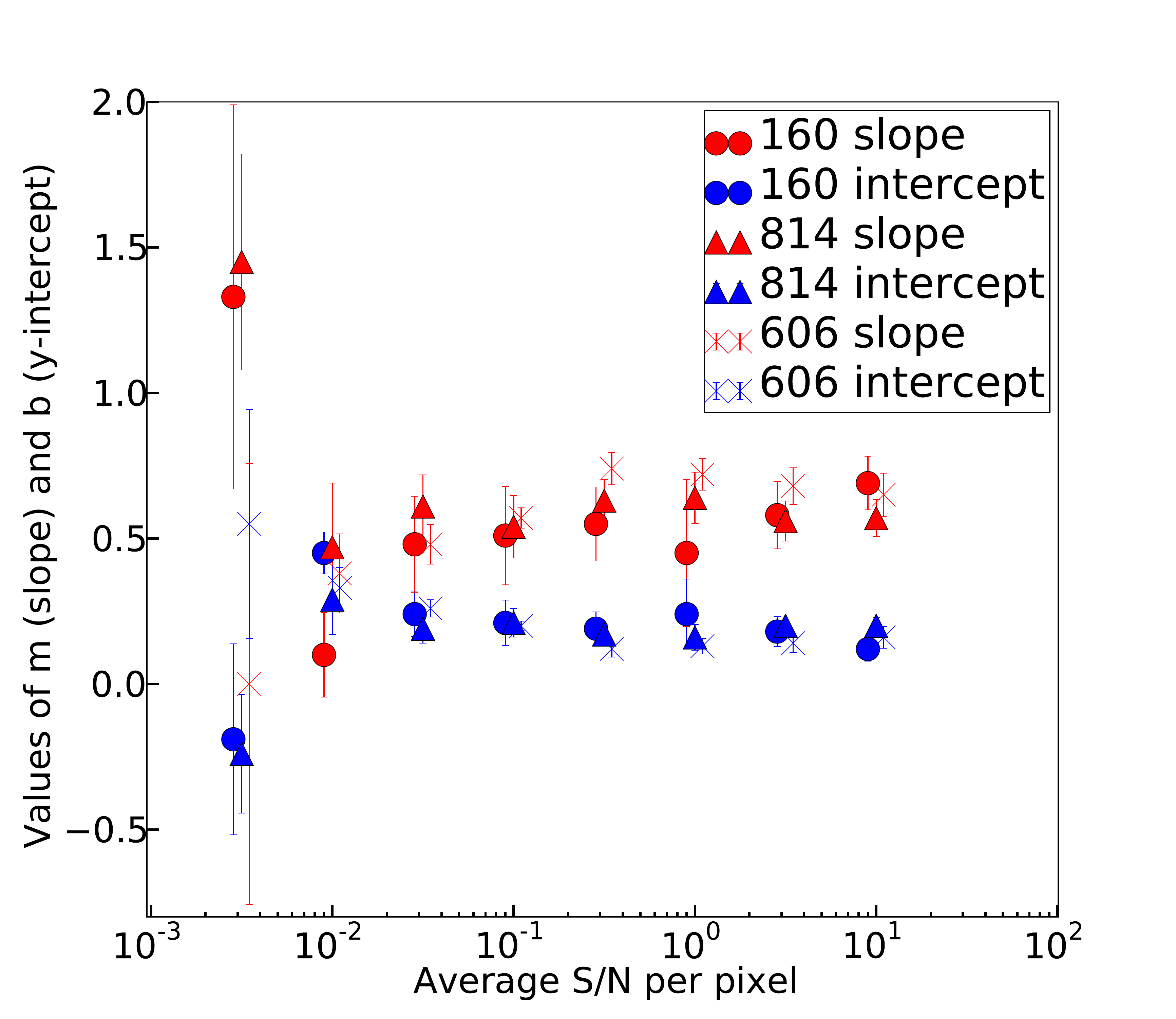}
\caption{Values of the slope and y intercepts of the best fitting lines for unlensed vs. lensed Gini coefficient plots like the one in Fig.~\ref{LGCvUGC}, but for values of average S/N per pixel between 10$^{-2.5}$ and 10.  Note that at each S/N level, a small offset has been artificially introduced for each filter, to aid in clarity by preventing points from overlapping with each other.  There is little change in these parameters until S/N per pixel drops below 10$^{-1.5}$}
    \label{fitparams}

\end{figure}

\begin{figure*}
\centering
\includegraphics[width=0.75\textwidth]{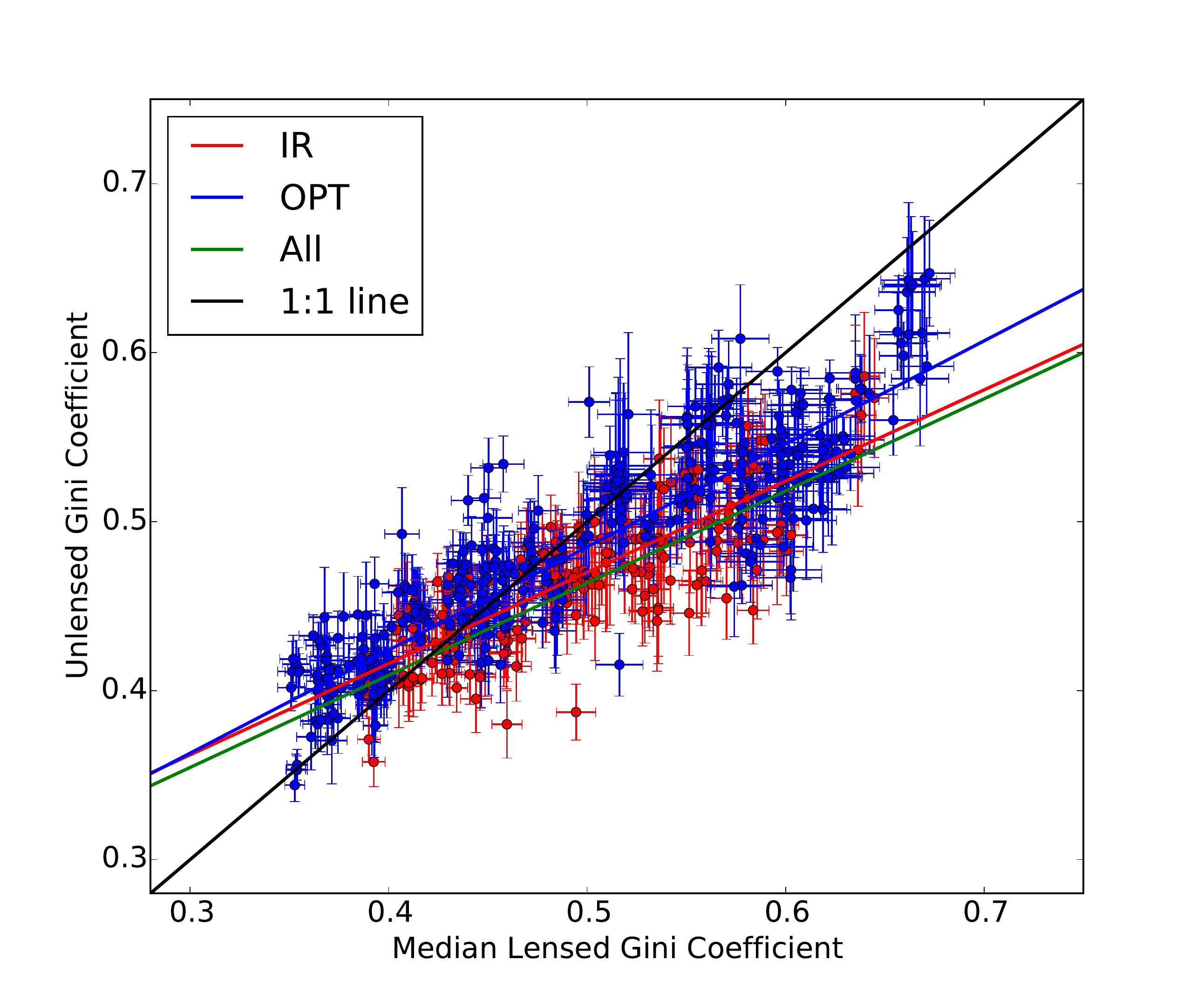}
\caption{Median lensed Gini coefficients plotted against unlensed Gini coefficients for all filters, with optical filters in blue and the IR filter in red.  Maximum likelihood lines are plotted for the optical filters (in blue) and the IR filters (in red).  All values of average S/N per pixel of 10$^{-1.5}$ and higher are used.  The parameters for these best fit lines are given in section 3.1.}
    \label{LGCvUGC_all}
\end{figure*}

For each source galaxy, in each S/N ratio bin and in each filter, the median Gini coefficient of all of the lensed images of that galaxy was plotted against its unlensed Gini coefficient.  The resulting plot for model realizations in the 0.1 S/N per pixel bin is shown in Fig.~\ref{LGCvUGC}.  Uncertainties in the median lensed Gini coefficients are simply the standard deviations in the distribution of Gini coefficients for lensed images of each source galaxy in the relevant filter.  The uncertainty in the unlensed Gini coefficient for each galaxy is calculated using the bootstrapping method described in \citet{Ab2003}.  Best fit lines for each filter were calculated using a maximum likelihood technique with weighting for each point determined by uncertainties.  While each filter has a slightly different slope and y-intercept, these parameters are relatively unchanged by the S/N per pixel values in the high S/N bins.  This relationship is shown in Fig.~\ref{fitparams} (note that a slight displacement along the S/N per pixel axis has been introduced for clarity).  Uncertainties were again determined by bootstrapping the selection of source galaxies used in the trendline analysis.  At most S/N levels higher than about 0.01/pixel, the best fitting parameters do not vary significantly with S/N.  Furthermore, the two optical ACS filters (which had very similar PSFs and whose PSFs were sharper than for the WFC3 IR filter) are indistinguishable from each other.  The slope of the best fit line for the F160W filter images tends to be shallower than for the other two filters regardless of S/N level until the noise becomes very high.  Therefore to characterize the relationship between the lensed and unlensed Gini coefficients, we fit one line to the optical (sharper PSF) data and another line to the IR (broader PSF) data.  In each case, the lines were fit to a dataset consisting of all points from all 6 S/N levels greater than 0.01 using a maximum likelihood technique with the significance of each point weighted according to the uncertainties in the lensed and unlensed Gini coefficients.  All points used in this analysis are shown in Fig.~\ref{LGCvUGC_all} along with the best fit lines for the IR sample, the optical sample, and the entire sample.  The following are the equations describing the best fit lines where UGC is the unlensed Gini coefficient and LGC is the lensed Gini coefficient:
\\
\\
IR: UGC = (0.54$\pm$0.04)$\times$LGC + 0.20$\pm$0.02
\\
OPTICAL: UGC = (0.61$\pm$0.02)$\times$LGC + 0.18$\pm$0.01
\\
ALL: UGC = (0.60$\pm$0.02)$\times$LGC + 0.18$\pm$0.01
\\

It is interesting to note that the increased spread in the lensed Gini coefficient relative to the unlensed Gini coefficients suggests that there is more spatial information available in the lensed images, as one might expect because of the additional magnification.  However, the fact that they are so similar to the unlensed Gini coefficients also suggests that most of the structure that determines the Gini coefficient is still visible at the resolution used for the unlensed galaxies.  This should not be surprising, however, since previous studies have shown that HST-like resolutions are sufficient to extract meaningful morphological information from the Gini coefficient of galaxies even up to $z\approx4$ \citep{Lotz06}.  This means that it should be possible to use the Gini coefficient to compare samples of strongly lensed galaxies to their unlensed counterparts at similar redshifts, which could allow questions of selection effects in strongly lensed samples to be better addressed.

It is also worth noting that both the lensed and unlensed values of the Gini coefficients typically extended to higher values in the bluer filters even though the Gini coefficients behaved similarly near the lower end of the range in all filters.  This relationship persisted even when the images originally taken in the F606W filter were convolved with a wider Gaussian to achieve a PSF similar to that of the F160W filter.  This suggests that the effect actually has an astrophysical interpretation and is not just the result of PSF convolution and pixelization.  If, for example, young and old stellar populations are spatially distributed with different uniformity in a particular galaxy, a mismatch between the Gini coefficients across these two filters would be expected.  If this were the case, then it indicates that comparing Gini coefficients measured in different filters may yield further information about the morphology and stellar structure of a galaxy.

\subsection{Gini Coefficients and the Effect of S/N Ratio}
\begin{figure}
\plotone{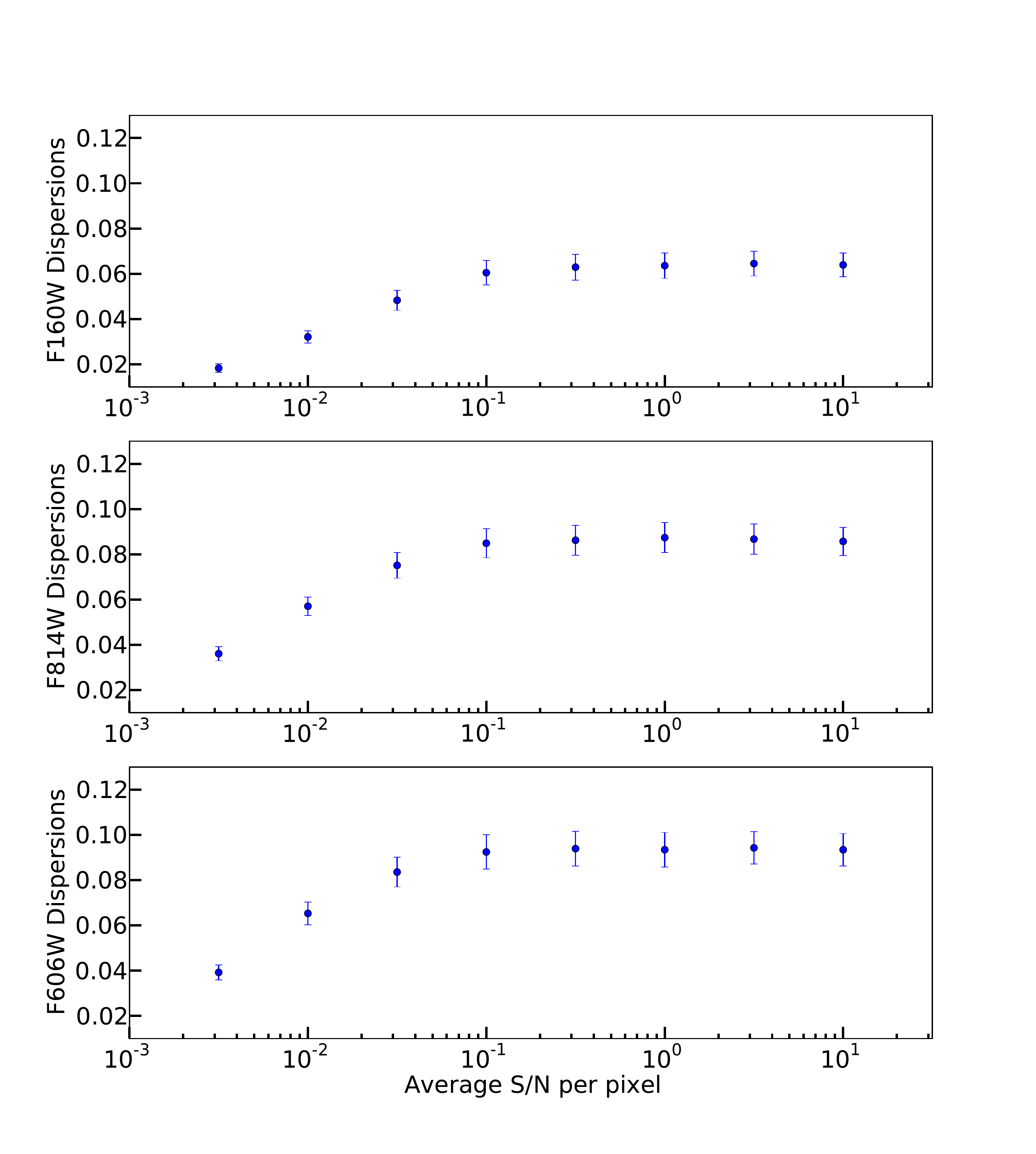}
\caption{The dispersion in the 33 median Gini coefficients at each S/N level.  As S/N drops, the Gini coefficients for each galaxy approach the same value, but at high S/N levels, the dispersion in the Gini coefficients of the 33 distinct galaxies is much higher.}
    \label{dispersions}
\end{figure}

We have seen in the previous section that the standard deviation of the Gini coefficient in different model realizations of the same lensed source galaxy is small relative to the overall dispersion of the median lensed Gini coefficients of the 33 different source galaxies.  This means that the Gini coefficient can often be used to distinguish between the lensed images of different source galaxies \citep{Florian2}.  However, it is easy to see that in the extreme case where the average S/N per pixel is very low, the noise will dominate the Gini coefficient measurement and the Gini coefficients of the 33 different galaxies will begin to converge.  The natural question to ask is: what is the minimum average S/N per pixel required for the conclusions of section 3.1 to hold?  To test this, we plot the dispersion of the 33 median Gini coefficients against average S/N per pixel.  That is, from all of the measured strongly-lensed images of each galaxy, we calculate the median Gini coefficient, and take the standard deviation of the 33 medians (one for each source galaxy).  When noise is not the dominant source of flux (i.e., when the galaxies' Gini coefficients are discernibly different), we should expect a high dispersion, but the dispersion should tend toward zero as the S/N level decreases.  This is borne out in Fig.~\ref{dispersions} (where uncertainties in the dispersion are calculated by bootstrapping).  It appears that the Gini coefficient is most informative at average S/N per pixel greater than or equal to about 0.1.

\subsection{Lensed Gini Coefficients of Multiply-Imaged Sources}
\begin{figure}
\plotone{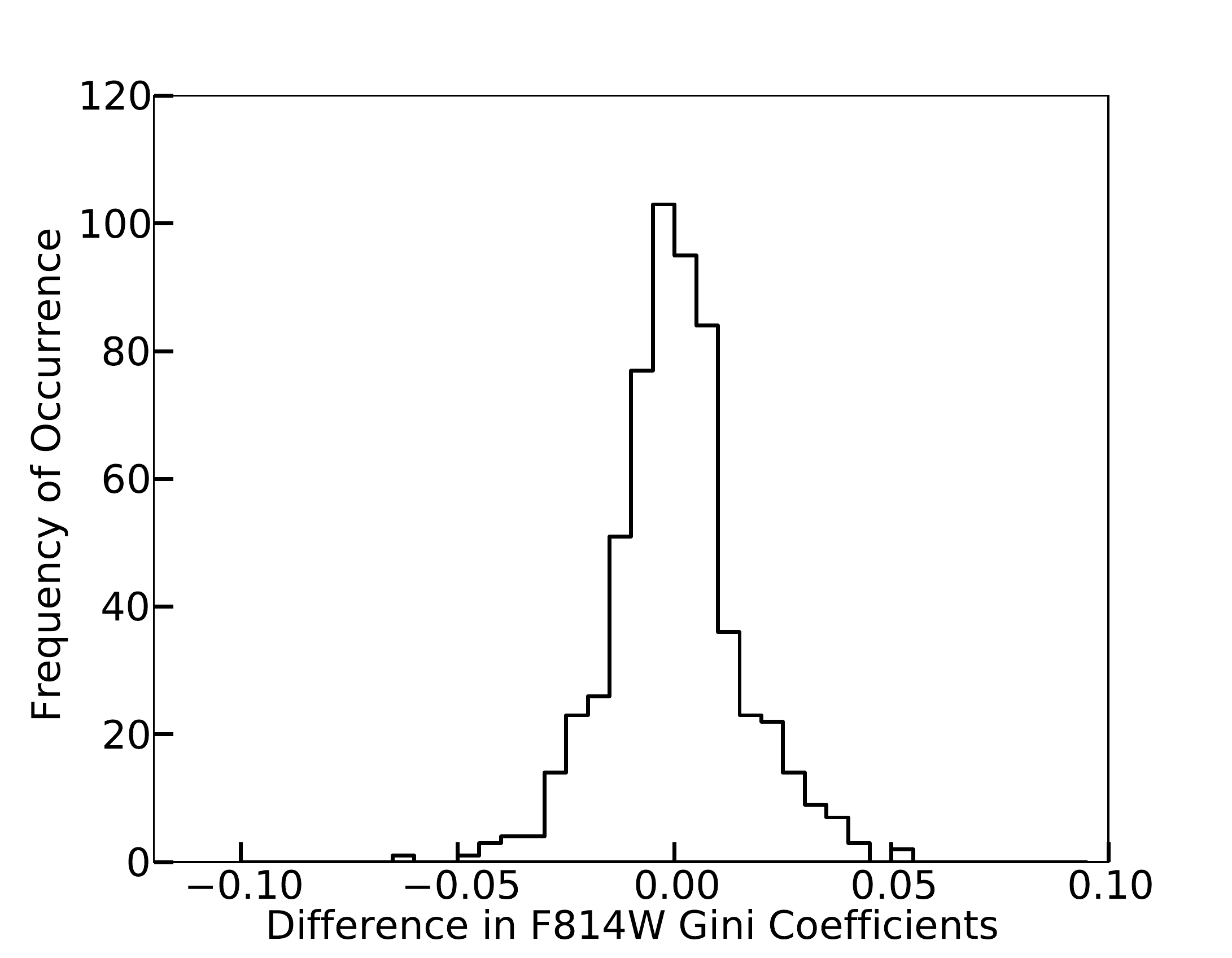}
\caption{The distribution of differences in F814W Gini coefficients between pairs of images from all lensing configurations in the sample for which there are multiple images produced.  The order of the pairs is selected randomly, leading to both negative and positive values.  The average S/N per pixel for arcs used in this figure was 0.1.  The narrowness of this distribution confirms that different images of the same galaxy should have very similar Gini coefficients even when strongly gravitationally lensed.}
    \label{imagePairs}

\end{figure}

One way to check the stability of the Gini coefficient under gravitational lensing using real observational data would be to compare the Gini coefficients of galaxies that have been lensed in such a way as to produce multiple images.  This can also be done with the simulated images.  We have taken pairs from multiple-image configurations and subtracted their Gini coefficients.  The histogram of the resulting distribution for the S/N = 0.01 per pixel subsample is shown in Fig.~\ref{imagePairs}.  Because the Gini coefficients were picked in random order, the distribution contains both negative and positive values and is, predictably, centered at zero.  However, examining the spread of these differences in Gini coefficient, we find that the standard deviation is only about 0.015 while the total range in Gini coefficients as seen in Fig.~\ref{LGCvUGC} runs from just under 0.4 to just over 0.6 (though these are medians--some individual images have gini coefficents that range closer to 0.3 or 0.7).  This shows that the Gini coefficients of different images of the same galaxy should be expected to be consistent with each other, and that any differences are small compared to the differences possible based on actual structural differences between two different galaxies.  Furthermore, it seems that differences in Gini coefficient due to changes in differential magnification from one realization to another are small.  However, it is worthwhile to more thoroughly investigate the effects of differential magnification, which we do in the following section.

\section{Elliptical Halos and the Effects of Higher Magnifications and Merging Images}
The case of the spherical halo, while allowing some shear and some magnification, does not permit certain classes of image configurations to form, including those with merging tangential arcs.  Furthermore, even in cases of complete images, differential magnifications can be much higher in the case of an elliptical halo, for example, than for a spherical one.  While we are encouraged by the results obtained in the spherical halo simulation, it would be prudent to further investigate the effects of these more exotic image configurations on the Gini coefficient.  There are a few types of halos that could be used to do this.  We could use a halo from a numerical simulation or a model from a Frontier Fields cluster \citep{FF1,FF2}, for instance.  The downsides of these options are that they introduce additional uncertainties due to, for example, density estimators or uncertain empirical deflection matrices.  Instead, since isothermal elliptical halos can produce merging images and high differential magnifications but are analytically defined, they allow better isolation of these particular effects.  Therefore, we have chosen to investigate these effects using an isothermal ellipsoid as the lensing mass.

\begin{figure*}
\centering
\includegraphics[width=1.00\textwidth]{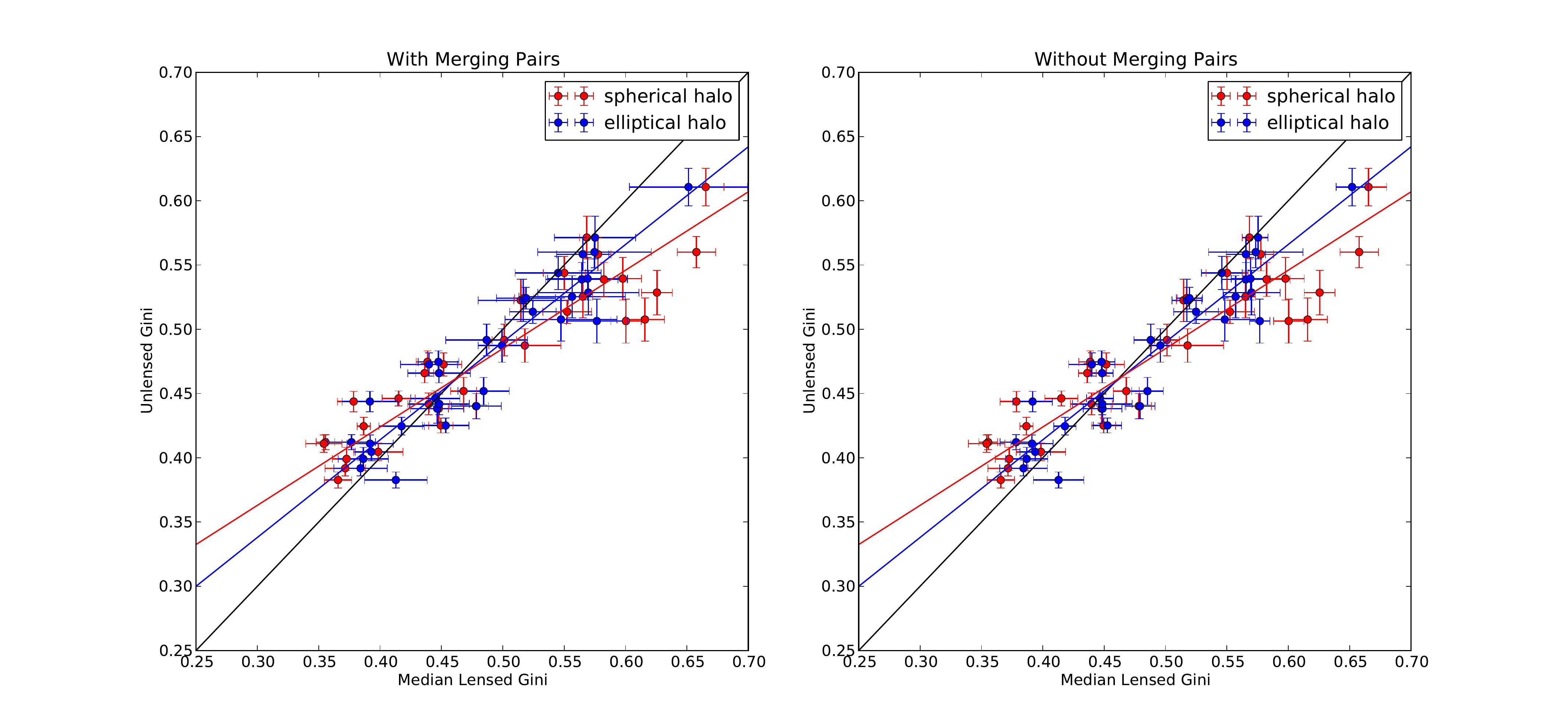}
\caption{The unlensed Gini coefficient as a function of the median lensed Gini coefficient of tangential arcs and counter images in the F606W filter for sources lensed by an elliptical halo (in blue) compared to sources lensed by a spherical halo (in red).  Uncertainties are calculated the same way as in Fig.~\ref{LGCvUGC}.  On the left, merging pairs are included in the sample.  On the right, they are excluded.  The best fit line is for the spherical case is plotted in red.  The best fit line for the elliptical case without merging pairs is plotted in blue.}
    \label{ell_ugc_lgc}
\end{figure*}

The same 33 source galaxies as before were used as input and were lensed by an isothermal ellipsoid with M$_{200}$ = 10$^{15}$ M$_{\odot}/h$, c = 5, $z_{l}$ = 0.2, $z_{s}$ = 0.1 a/b = 0.8.  Since sky noise had little effect when the S/N was at or above 0.01/pixel, only one S/N bin was used (S/N=0.1).  Images were visually categorized by whether they were tangential/counter images, radial images, or central images, as well as by whether they were merging images.  Apertures were created using the same prescription as in the spherical halo case.  When images were merging, there was no attempt to break the arc into component partial images since in real observational data, it would be unclear where the breaks should be.  The relationship between lensed and unlensed Gini coefficients is shown in Fig.~\ref{ell_ugc_lgc}, the analogue of Fig.~\ref{LGCvUGC} except, for simplicity, only the F606W images are included.  For comparison, the same values from the spherical halo case are also plotted.  It is clear that there is increased scatter in the ellitpical case.  This is largely due to contamination from high magnification merging pairs where the source galaxies cross a caustic and are not fully mapped into the image plane.  Instead, small fractions of these source galaxies are imaged twice, while other parts of them are not visible in the image plane at all.  When this happens, the measured Gini coefficient is not representative of the entire source galaxy, but only some small but doubly-imaged part of it.  With such points removed, the typical scatter is more in line with the spherical case.  The best fitting parameters for the line are (for the scatterplot excluding contamination from merging pairs):
\\
\\
UGC = (0.76$\pm$0.05)$\times$LGC + 0.11$\pm$0.02
\\
\\

\begin{figure*}
\centering
\includegraphics[width=0.75\textwidth]{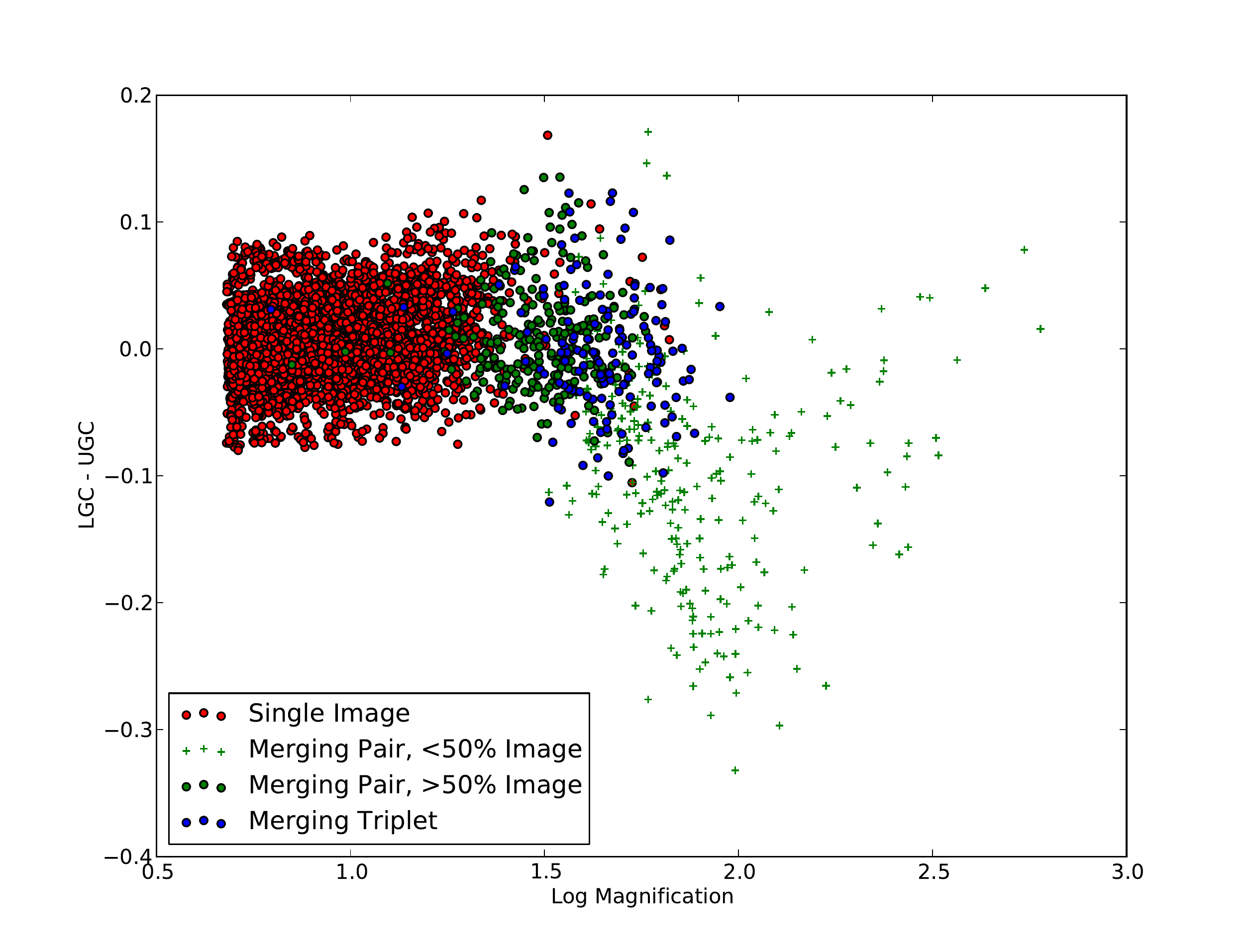}
\caption{The difference between the lensed and unlensed Gini coefficients of tangential arcs and counter images plotted against the log of the magnification.  Different colors correspond to isolated single images, merging pairs, and merging triplets, while circular points are for images where at least 50\% of the source image is visible in the lensed images and crosses when there less than 50\% is visible.  Only merging pairs have less than 50\% of the source image visible in the lensed image.  Lensed Gini coefficients deviate from the unlensed Gini coefficients by less than about 0.08 across a wide range of magnifications for the single images and the merging triplets.  For merging pairs where only a small fraction of the source image is visible, however, the lensed Gini coefficient can deviate significantly from the unlensed Gini coefficient.  Care must be taken in the interpretation of Gini coefficients for merging pairs of images.}
    \label{magnification_effects}
\end{figure*}

\begin{figure*}
\centering
\includegraphics[width=1.00\textwidth]{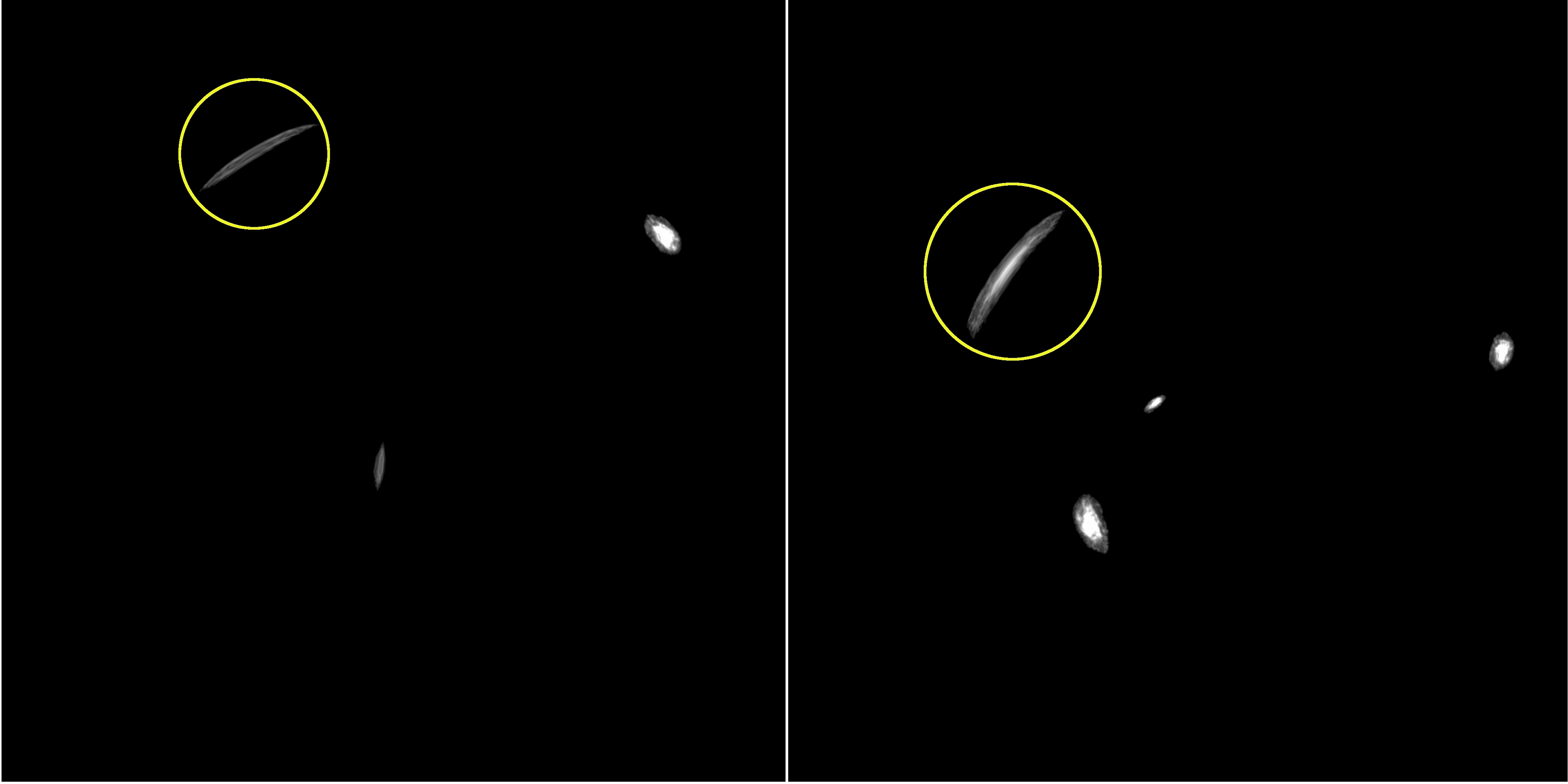}
\caption{On the left: A lensing configuration with a merging pair of tangential arcs (circled) where the resulting image does not inlude the center of the source galaxy.  The Gini coefficient of this merging pair is less than the unlensed Gini coefficient for the source galaxy by more than 0.1.  On the right: A lensing configuration with a merging pair of tangential arcs (circled) where the center of the source galaxy is visible.  The Gini coefficient of this merging pair agrees with the unlensed gini coefficient within about 0.01.  For both the left and right image, the same source galaxy was used.}
    \label{good_bad_mergers}
\end{figure*}

To illustrate this effect, we have plotted, in Fig.~\ref{magnification_effects}, the difference between the lensed and unlensed Gini coefficients as a function of magnification for all tangential and counter images.  The points are colored based on whether they represent values from individual isolated images, merging pairs, or merging triplets.  No aperture was applied in the lensed image when calculating magnifications (i.e., all nonzero points in the noise-free simulated image were included).  The figure demonstrates several important points.  One is that except for some cases of merging pairs, where the lensed Gini coefficient is typically lower than the unlensed one, there is no noticeable bias in the Gini coefficient with increasing magnification, nor is there a clear difference in scatter across magnifications that are well-represented in our simulated data.  Another is that high magnification merging pairs tend to have their Gini coefficients lowered by lensing, but isolated images and merging triplets do not show this bias.  This is easy to understand in light of the example images in Fig.~\ref{good_bad_mergers}.  On the left, the circled image is of a merging pair where only a small part of the galaxy is visible in the lensed image.  This small, and fairly uniformly faint portion of the galaxy is imaged twice, but the brighter central region (the region that will contribute the most to increasing the Gini coefficient) is not visible.  In this case, the observed Gini coefficient is very low.  However, the arc would also have a very low surface brightness and may not be observable in real data.  And even if it was, no other photometric measurement would be able to tie it definitively to other members of its image family.  On the right, the circled image is of a merging pair where the central region of the source galaxy appears in the observed image.  As a result, its Gini coefficient hardly differs from its unlensed counterpart.

The results of this section reaffirm the consistency of the Gini coefficient as a morphological metric in lensed galaxies across a wide range of magnfications and image classes.  They do, however, also draw attention to an important caveat.  Care must be taken when interpreting the Gini coefficient of merging pairs of images.  If a large portion of the source galaxy is visible in each of the merging images, it is more likely that the Gini coefficient will be consistent with the unlensed source and with other lensed images of the same source.  If only part of the image is visible, the Gini coefficient is unlikely to be a reliable metric.  Interestingly, in the case of merging triplets, enough of the source galaxy is almost always observable in the lensed images that the Gini coefficient does not change significantly from its unlensed value.

\section{Conclusions}
At HST resolutions, the Gini coefficients of galaxies at high redshifts are well-preserved under strong gravitational lensing.  However, because of PSF effects and pixelization, the conservation is not perfect.  Fortunately, these effects are relatively minor, and can easily be calibrated out.  A small deviation from the Gini coefficient in the unlensed frame exists, manifesting as an increased slope and scatter in the plot of unlensed vs lensed Gini coefficients in the studied range of about 0.2 to about 0.7, but scatter off of the trendlines and a slope of greater than 1 is not unreasonable because of the extra spatial information captured in the strongly lensed images.  However, the differences in the trendlines for the IR and optical subsamples and between those subsamples and the entire ensemble are only marginal.  There also does not appear to be a significant difference in lensed Gini coefficients in two images of the same source based on magnification.  However, the fraction of the source galaxy that is visible in the image plane does make an important difference.  The Gini coefficient is not preserved when the image fraction is less than about 0.5-0.6.

Since a relationship between unlensed and lensed Gini coefficients exists, and because unlensed Gini coefficients continue to carry morphological information out to redshift 4 \citep{Lotz06}, we can begin to contextualize the existing samples of lensed galaxies like those in the Sloan Giant Arcs Survey (SGAS) sample (e.g., \citealp{Bayliss11,Bayliss14,Sharon14}) or the Hubble Frontier Fields \citep{FF1,FF2} with their unlensed counterparts imaged across redshifts in deep field surveys like the Hubble Ultra Deep Field, CANDELS, and others.  However, because the aperture definitions are different, a comparison of Gini coefficients between lensed and unlensed samples will require recalculation of the Gini coefficients of unlensed galaxies using the aperture definition in section 2 of this paper.  In addition to this application, the preservation of the Gini coefficient in the image plane provides morphological information that can be used as a constraint for identifying image families for the purposes of lens modeling \citep{Florian2} in, for example, the Frontier Fields clusters.

Most importantly, we have shown that the Gini coefficient is indeed a meaningful measurement of galaxy morphology that can be conducted in the image plane---no source plane reconstruction is necessary, which means that no lens model is needed.  Lens modeling and source plane reconstruction are processes that are time and resource intensive (including both astronomers and telescopes), so measurements like the Gini coefficient, which can be made in the image plane and in only one filter, provide workarounds for one of the most substantial bottlenecks in the process of understanding the morphology of strongly lensed high redshift galaxies.  While some applications will still require lens modeling and source plane reconstruction, there is certainly morphological information that can be gleaned from strongly lensed images without these extra steps.

\acknowledgments
Argonne National Laboratory's work was supported under the U.S. Department of Energy contract DE-AC02-06CH11357.

This work was supported in part by the Kavli Institute for Cosmological Physics at the University of Chicago through grant NSF PHY-1125897 and an endowment from the Kavli Foundation and its founder Fred Kavli, and by the Strategic Collaborative Initiative administered by the University of Chicago's Office of the Vice President for Research and for National Laboratories.

We would also like to thank the anonymous referee for the thorough and constructive review of our original manuscript that allowed us to greatly improve the quality of this paper.

\clearpage


\begin{thebibliography}{}
\bibitem[Abraham et al.(1996)]{Ab1996} Abraham, R. G., Tanvir, N. R., Santiago, B.X., et al. 1996, \mnras, 271, 47
\bibitem[Abraham et al.(2003)]{Ab2003} Abraham, R. G., van den Bergh, S., \& Nair, P.  2003, \apj, 588, 218
\bibitem[Bayliss et al.(2014)]{Bayliss14} Bayliss, M. B., Rigby, J. R., Sharon, K., et al. 2014, \apj, 790, 144
\bibitem[Bayliss et al.(2011)]{Bayliss11} Bayliss, M. B., Gladders, M. D., Masamune, O., et al. 2014, \apj, 790, 144
\bibitem[Bertin \& Arnouts(1996)]{SExtractor} Bertin, E. \& Arnouts, S. 1996, A\&AS, 117, 393
\bibitem[Blanton et al.(2001)]{Blant2001} Blanton, M. R. et al. 2001, \aj, 121, 2358
\bibitem[Conselice(2003)]{Con2003} Conselice, C. J. 2003, \apjs, 147, 1
\bibitem[Dressler(1980)]{Dress80} Dressler, A. 1980, \apj, 236, 351
\bibitem[Florian et al.(2016)]{Florian2} Florian, M., Gladders, M., Li, N., Sharon, K. 2016, \apjl, 816, 23
\bibitem[Galametz(2013)]{CANDELS_UDS} Galametz, A., Grazian, A., Fontana, A., et al. 2013, \apj, 206, 10
\bibitem[Grogin et al.(2011)]{Grogin11} Grogin, N. A., Kocevski, D., Faber, S. et al. 2011, ApJS, 197, 37
\bibitem[Jones et al.(2010)]{Jones10} Jones, T.A., Swinbank, A.M., Ellis, R.S., Richard, J., Stark, D.P. 2010, \mnras, 404, 124
\bibitem[Jones et al.(2013)]{Jones13} Jones, T., Ellis, R.S., Richard, J., Jullo, E. 2013, \apj, 765, 48
\bibitem[Koekemoer et al.(2011)]{Koek11} Koekemoer, A. M., Faber, S., Ferguson, H. et al. 2011, ApJS, 197, 36
\bibitem[Koekemoer et al.(2014)]{FF2} Koekemoer, A. M., Avila, R. J., Hammer, D., et al. 2014, AAS, 223, 254.02
\bibitem[Li et al.(2016)]{Li2015} Li, N., et al. 2016 (submitted)
\bibitem[Lisker(2008)]{Lisk2008} Lisker, T. 2008, \apjs, 179, 319
\bibitem[Lotz et al.(2004)]{Lotz04} Lotz, J.M., Primack, J., Madau, P. 2004, \apj, 128, 163
\bibitem[Lotz et al.(2006)]{Lotz06} Lotz, J.M., Madau, P., Giavalisco, M., Primack, J., Ferguson, H.C. 2006, \apj, 636, 592
\bibitem[Lotz et al.(2014)]{FF1} Lotz, J. M., Bountain, M., Grogin, N. A., et al. 2014, AAS, 223, 254.01
\bibitem[Papovich et al.(2003)]{Pap2003} Papovich, C., Giavalisco, M., Dickinson, M., Conselice, C. J., Ferguson, H. C. 2003, \apj, 598, 827
\bibitem[Lefor et al.(2013)]{Lef13} Lefor, A.T., Futamase, T., Akhlaghi, M. 2013, NewAR, 57, 1, 1
\bibitem[Sharon et al.(2014)]{Sharon14} Sharon, K., Gladders, M. D., Rigby, J. R., et al. 2014 \apj, 790, 50
\bibitem[Swinbank et al.(2009)]{Swin09} Swinbank, A.M., Webb, T.M., Richard, J., et al. 2009, \mnras, 400, 1121
\bibitem[Willett et al.(2015)]{Will15} Willet, K. W., Schawinski, K., Simmons, B. D., et al. 2016, \mnras, 449, 820
\bibitem[Wuyts et al.(2014)]{Wuyts14} Wuyts, E., Ribgy, J.R., Gladders, M.D., Sharon, K. 2014, \apj, 781, 61
\bibitem[Yasuda et al.(2001)]{Yas2001} Yasuda, N. et al. 2001, \aj, 122, 1104
\end{thebibliography}
\end{document}